\documentstyle[11pt,epsfig]{article} \textheight 700pt \textwidth
480pt \oddsidemargin 0pt \voffset -2.5cm
\title{\bf Classical tests in brane gravity}
\author{S. Jalalzadeh\thanks{email: s-jalalzadeh@sbu.ac.ir},\,\,\ M. Mehrnia\thanks{email: m.mehrnia@sbu.ac.ir}\,\
and H. R. Sepangi\thanks{email: hr-sepangi@sbu.ac.ir}
\\{\small  \emph{Department of Physics, Shahid
Beheshti University, Evin, Tehran 19839, Iran}}}
\begin{document}
\maketitle
\begin{abstract}
The vacuum solutions in brane gravity differ from those in $4D$ by a
number of additional terms and reduce to the familiar Schwarzschild
metric at small distances. We study the possible roles that such
terms  may play in the precession of planetary orbits,  bending of
light, radar retardation and  the anomaly in mean motion of test
bodies. Using the available data from Solar System experiments, we
determine the range of the free parameters associated with the
linear term in the metric. The best results come from the anomalies
in the mean motion of planets. Such studies should shed some light
on the origin of dark energy via the solar system tests.
\vspace{5mm}\noindent
\end{abstract}
\section{Introduction} In the past two decades,
various attempts have been made to understand the nature of the
so-called dark energy and dark matter which would be required to
explain many observational data. For a review, see for example
\cite{1,2,3,4}. Deviation of the galactic rotation curves from
Newtonian gravity which occurs at distances larger than the Solar
System scales \cite{5}, the velocity of galaxies in clusters and the
bending of light rays from galaxies and clusters \cite{6} are the
best evidence for the existence of copious amounts of dark matter.
Apart from the efforts made by a number of authors to account for
dark matter by considering such elusive objects as massive
neutrinos, axions and the Weakly Interacting Massive Particles
(WIMPS) which is predicted by supersymmetric theories, the nature of
dark matter is still unresolved. An interesting attempt to explain
the galactic rotation curves was made in \cite{7} where the author
considers dark matter as a galactic phenomena and proposes what is
now known as the Modified Newtonian Dynamics (MOND) at low
accelerations. However, since MOND is a non-relativistic theory, it
cannot predict any relativistic phenomena such as the bending of
light, the precession of planetary orbits etc. Solar System
experiments have been increasingly relied upon to investigate the
integrity of the foundation of general relativity in recent years,
see for example \cite{75}-\cite{85}.

In the recent past, models incorporating extra dimensions  have
become the focus of attention for investigating the nature of
problems mentioned above. In these theories one considers a four
dimensional world (brane) embedded in a higher dimensional manifold
(bulk) through which only gravity can propagate. Ordinary matter is
confined to the brane and cannot propagate through the bulk. The
confinement is achieved, through the imposition of  $Z_{2}$ symmetry
and use of the Israel junction conditions which relates the
extrinsic curvature of the brane to the energy momentum of the
matter.  This method has predominantly been used in theories with
one extra dimension. If the number of extra dimensions exceeds one,
no reliable method for confining matter to the brane exists. This is
so since the requirement to define junction conditions is the
existence of a boundary (brane) which cannot be defined if the
number of extra dimensions is more than one.  For example, a
boundary surface in a $3D$ space is a surface with one less
dimension whereas a line in the same space cannot be considered as
its boundary. On the back of such concerns, model theories have been
proposed where matter is confined to the brane through the action of
a confining potential, without the use of any junction condition or
$Z_2$ symmetry \cite{8}. In \cite{9} the authors used the confining
potential approach to study a brane-world embedded in a
$m$-dimensional bulk. The field equations obtained on the brane
contained an extra term which was identified with the X-cold dark
matter. The same methodology was used in \cite{10} to find the
spherically symmetric vacuum solutions of the field equations on the
brane. These solutions were shown to account for the accelerated
expansion of the universe and offered an explanation for the galaxy
rotation curves.

In this paper, we focus attention on the consequences of the
spherically symmetric vacuum solutions mentioned above when
considering such questions as the precession of planetary orbits,
the deflection of light rays in the Solar System, the time delay of
signals in the Solar System and the mean motion of test bodies. In
doing so, we obtain constraints on the free parameters appearing in
the metric. This should hep us to accounting for the origin of dark
energy via Solar System tests.
\section{The model}
Let us start by presenting the model used in our calculations. We
only state the results and refer the reader to \cite{9,10} for a
detailed derivation of these results.

As was mentioned in the introduction, the brane-world model we
invoke here differs from the usual Randall-Sundrum type in that no
junction conditions or $Z_2$ symmetry is used. One thus starts with
the usual setup in which a $4D$ brane is embedded in a 5 or, in
general, $n$-dimensional bulk. Assuming that the brane is devoid of
matter with no cosmological constant and that the bulk space has
constant curvature, one arrives at the following equations
\cite{10,1010}
\begin{eqnarray}
G_{\mu\nu}= Q_{\mu\nu},\label{6}
\end{eqnarray}
where
\begin{eqnarray}
Q_{\mu\nu}=\left(K^{\rho}_{\,\,\,\,\mu
}K_{\rho\nu}-KK_{\mu\nu}\right)-\frac{1}{2}
\left(K_{\alpha\beta}K^{\alpha\beta}-K^2\right)g_{\mu\nu}.\label{7}
\end{eqnarray}
We note that $Q_{\mu\nu}$ is an independently conserved quantity,
that is
\begin{eqnarray}
Q^{\mu\nu}_{\,\,\,\,;\mu}=0,\label{8}
\end{eqnarray}
so that equation (\ref{6}) satisfies the covariant conservation
law. Equation (\ref{6}) is the starting point from which a class
of solutions were found in \cite{1010}, representing a black hole.
Thus, starting with the metric
\begin{eqnarray}
ds^2=-e^{\mu(r)}dt^2+e^{\nu(r)}dr^2+r^2\left(d\theta^2+\sin^2\theta
d\varphi^2\right),
\end{eqnarray}
the following solution satisfies equations (\ref{6}) and (\ref{8})
\begin{eqnarray}\label{2,12}
e^{\mu}=e^{-\nu}=1-\frac{C}{r}-\alpha^2r^2-2\gamma r-\beta^2.
\end{eqnarray}
This solution represents a black hole which we shall consider in
the next section and use to analyze the behavior of a test
particle in such a space-time.
\section{The geodesic equations of motion}
The induced line element of the vacuum $4D$ space-time is
\begin{eqnarray}\label{3,13}
ds^2=A(r)dt^2-\frac{dr^2}{A(r)}-r^2 d\Omega^2,
\end{eqnarray}
where
\begin{eqnarray}\label{14}
A(r)=1-\frac{2m}{r}-\alpha^2r^2-2\gamma r-\beta^2,
\end{eqnarray}
with $\gamma\equiv\alpha\beta$,  $m$ is the mass of the central
object, $\alpha$ and $\beta$ are constants and we have adopted the
relativistic units, $c=G=1$. In this space-time, the geodesic
equations of motion for a test particle are the Euler equations
resulting from the lagrangian
\begin{eqnarray}\label{3,15}
L=\frac{1}{2}\left[A(r)\dot{t}^2-\frac{\dot{r}^2}{A(r)}-
r^2\left(\dot{\theta}^2+\sin^2\theta\dot{\varphi}^2\right)\right],
\end{eqnarray}
where a dot denotes differentiation with respect to the affine
parameter. Without loss of generality, let us consider the
equatorial plane, $\theta=\frac{\pi}{2}$, for a test particle. The
lagrangian then becomes
\begin{eqnarray}\label{3,16}
L=\frac{1}{2}\left[A(r)\dot{t}^2-\frac{\dot{r}^2}{A(r)}-r^2\dot{\varphi}^2\right].\label{eq1}
\end{eqnarray}
Now, using the Euler-Lagrange equations, we obtain
\begin{eqnarray}\label{17}
\dot{t}=\frac{E}{A(r)},
\end{eqnarray}
\begin{eqnarray}\label{3,18}
\dot{\varphi}=\frac{J}{r^2},
\end{eqnarray}
with $J$ and $E$ being constants. Along the orbit for a test
particle $2L=1$ and for photons $2L=0$. Now, using equations
(\ref{17}), (\ref{3,18}) in equation (\ref{eq1}) we obtain
\begin{eqnarray}\label{3,19}
\frac{1}{A(r)}\left[E^2-\dot{r}^2\right]-\frac{J^2}{r^2}=C,
\end{eqnarray}
where $C=1$ and $C=0$ represent the massive and massless particles
respectively.
\section{Precession of planetary orbits}
To find equation of motion of a massive particle in this space-time
we use  equation (\ref{3,19}) $(C=1)$ and obtain
\begin{eqnarray}\label{4,20}
u'^2+\left(1-\beta^2\right)u^2=\frac{E^2+\beta^2-1}{J^2}+\frac{2mu}{J^2}+2\gamma
u+2mu^3+\frac{\alpha^2}{J^2u^2}+\frac{2\gamma}{J^2u}+\alpha^2,
\end{eqnarray}
where $u=\frac{1}{r} $ and a prime denotes differentiation with
respect to $\varphi$. Now by differentiating equation (\ref{4,20})
with respect to $\varphi$ we find the differential equation of the
motion for a massive particle as
\begin{eqnarray}\label{4,21}
\frac{d^2u}{d\varphi^2}+\left(1-\beta^2\right)u=
\frac{m}{J^2}+3mu^2+\gamma-\frac{\alpha^2}{J^2u^3}-\frac{\gamma}{J^2u^2}.
\end{eqnarray}
In order to solve this equation we consider $3mu^2$,
$\frac{\alpha^2}{J^2u^3}$ and $\frac{\gamma}{J^2u^2}$ as
perturbative terms, since these terms are much smaller than
$\frac{m}{J^2}$. This can be seen by noting that, for example,
$3mu^2/(m/J^2)=(3/r^2)(r^2\dot{\varphi})^2\approx
3(rd\varphi/dt)^2/c^2\approx 3v^2/c^2\approx 7.7\times 10^{-8}$ for
Mercury. To first order then, this equation has a solution of the
form
\begin{eqnarray}\label{4,22}
u\simeq
\frac{1}{P}\left[1+e\cos\left(1-\frac{1}{1-\beta^2}\left(\frac{3\alpha^2P^4}{2J^2}+\frac{\gamma
P^3}{J^2}+\frac{3m}{P}\right)\right)\left(1-\beta^2\right)^{1/2}\varphi\right],
\end{eqnarray}
where
\begin{eqnarray}\label{4,23}
P=\frac{1-\beta^2}{\gamma+\frac{m}{J^2}},
\end{eqnarray}
and $e$ is the eccentricity. From equation (\ref{4,22}) we find that
$r$ is a periodic function of $\varphi$ with period
\begin{eqnarray}\label{4,24}
2\pi\left[\left(1-\beta^{2}\right)^{1/2}-\frac{1}{\left(1-\beta^{2}\right)^{1/2}}\left(\frac{3\alpha^2P^4}{2J^2}+\frac{\gamma
P^3}{J^2}+\frac{3m}{P}\right)\right]^{-1}>2\pi.
\end{eqnarray}
Since $\beta^{2}, \gamma\ll 1$, to first order approximation, the
perihelion anomaly $\Delta\varphi$ of the particle after one
revolution is found to be
\begin{eqnarray}\label{4,25}
\Delta\varphi\simeq2\pi\left[\frac{3m^{2}}{J^2}+\gamma\left(\frac{J^{4}}{m^{3}}+3m\right)+\frac{3J^{6}\alpha^{2}}{2m^{4}}+\left(\frac{1}{2}+\frac{9m^{2}}{2J^{2}}\right)\beta^{2}\right].
\end{eqnarray}
For $\gamma>0$, we find that the perihelion anomaly will increase
compared to what one has for the Schwarzschild space-time
\cite{11,12,1416}. In the case $\gamma<0$, depending on the
parameters representing the particle, the perihelion anomaly may
increase or decreases compared to what is predicted by the
Schwarzschild metric.
\section{Deflection of light rays}
In this section we investigate the deflection of light in a
space-time represented by the metric above in two cases, $\gamma<0$
and $\gamma>0$. We assume that both the observer and the light
source are far from the compact object whose presence would deflect
the light rays. The light emitted from the source approaches the
compact object, reaching the observer at infinity. Using equation
(\ref{3,19}) for light $(C=0)$ we obtain
\begin{eqnarray}\label{26}
u'^2+\left(1-\beta^2\right)u^2=\frac{E^2}{J^2}+2mu^3+\alpha^2+2\gamma
u.
\end{eqnarray}
Here, a prime denotes differentiation with respect to $\varphi$,
$u=\frac{1}{r}$ and $\varphi$ is the usual angle representing the
deflection. If we differentiate the last equation with respect to
$\varphi$ again we find
\begin{eqnarray}\label{27}
\frac{d^2u}{d\varphi^2}+\left(1-\beta^2\right)u=3mu^2+\gamma.
\end{eqnarray}
This equation with $\beta=0$ represents the deflection of light rays
in Schwarzschild space-time. We also note that
$\frac{3mu^2}{u}=3\frac{R_{s}}{r} \leq\frac{R_{s}}{R_{c}}$ is small
where $R_{s}$ and $R_{c}$ are the Schwarzschild and compact object
radii respectively. For example, for a compact object like the Sun
this is of the order of $10^{-6}$. We therefore have, retaining only
the first term on the right hand side of equation (\ref{27})
\begin{eqnarray}\label{28}
u_{1}=\frac{3m}{2R^2\left(1-\beta^2\right)}\left[1+\frac{1}{3}\cos2\left(1-\beta^2\right)^{1/2}\varphi\right].
\end{eqnarray}
Equation (\ref{27}) has also an exact solution if we only retain the
second term on the right hand side. Therefore, the general solution
is given by
\begin{eqnarray}\label{29}
u=\frac{\gamma}{1-\beta^2}+\frac{1}{R}\sin(1-\beta^2)^{1/2}\varphi+
\frac{3m}{2R^2\left(1-\beta^2\right)}\left[1+\frac{1}{3}\cos2(1-\beta^2)^{1/2}\varphi\right],
\end{eqnarray}
where $R$ is the shortest distance between the object and the
undeflected path of the light rays in flat space
\begin{eqnarray}\label{5,30}
R=r\sin\varphi.
\end{eqnarray}
At large distances, $r\rightarrow\infty$,
$\varphi\rightarrow\varphi_{\infty}\ll1$, and from equation
(\ref{29}) we obtain
\begin{eqnarray}\label{31}
\varphi_{\infty}=-\frac{R}{\left(1-\beta^2\right)^{3/2}}\left[\frac{2m}{R^2}+\gamma\right].
\end{eqnarray}
As can be seen, the total deflection is twice that of the absolute
value of (\ref{31})
\begin{eqnarray}\label{32}
\delta=\frac{1}{\left(1-\beta^2\right)^{3/2}}\frac{4m}{R}+2\gamma\frac{R}{\left(1-\beta^2\right)^{3/2}}.
\end{eqnarray}
The first term in (\ref{32}) with $\beta=0$ is the deflection angle
for the Schwarzschild metric \cite{11,12,1416} and the second term
is the result of the modification of the model. As can be seen, the
proper potential increases (decreases) for the bending angle if we
choose $\gamma>0$ ($\gamma<0$) respectively.
\section{Radar retardation}
We now calculate the time it takes for a photon to travel from one
point to another in the gravitational field of a central object. The
equation which we shall use here is (\ref{3,19}) with $C=0$. At the
point of the closest approach $r=r_{0}$, we have $dr/dt=0$. Equation
(\ref{3,19}) then yields
\begin{eqnarray}\label{33}
\frac{J^2}{E^2}=\frac{r_{0}^2}{A(r_{0})},
\end{eqnarray}
where $A(r)$ is given by equation (\ref{14}). Using equations
(\ref{17}), (\ref{3,19}) and (\ref{33}) we get the time taken by the
photons traversing the distance from $r_0$ to $r$
\begin{eqnarray}\label{6,34}
t=\int_{r_{0}}^{r}\frac{dr}{A(r)\left[1-\left(\frac{r_{0}}{r}\right)^2\frac{A(r)}{A(r_{0})}\right]^{1/2}}.
\end{eqnarray}
After substituting $A(r)$ in the above integral, we expand the
integrand to first order in  $\frac{m}{r}$, $\gamma r$ and
$\alpha^2r^2$ which are much less than unity and find
\begin{eqnarray}\label{6,35}
t\simeq\int_{r_{0}}^{r}\frac{1+\frac{m}{r}\left(2+\frac{r_{0}}{r+r_{0}}\right)+
\alpha^2\left(r^2-\frac{1}{2}{r_{0}}^2\right)+
\gamma\left(2r-\frac{{r_{0}}^2}{r+r_{0}}\right)+\beta^2}{\left[1-\left(\frac{r_{0}}{r}\right)^2\right]^{1/2}}dr.
\end{eqnarray}
This integral can be evaluated to give
\begin{eqnarray}\label{6,36}
t&\simeq&\left[1+\beta^2+\alpha^2\left(\frac{r^2}{3}+\frac{{r_{0}}^2}{6}\right)+
\gamma\left(\frac{r^2+{r_{0}}^2+rr_{0}}{r+r_{0}}\right)\right]
\sqrt{r^2-{r_{0}}^2}\nonumber\\&+&2m\ln\left(\frac{r+\sqrt{r^2-{r_{0}}^2}}{r_{0}}\right)+m\left(\frac{r-r_{0}}{r+r_{0}}\right)^{1/2}.
\end{eqnarray}
For $\alpha=\beta=m=0$, we get the time traveled by photons in a
straight line between $r$ and $r_{0}$ in a Minkowski space-time. For
$\alpha=\beta=0$ we get the Schwarzschild metric result
\cite{11,12,1416} and for $\gamma>0$ we find an increase in the time
delay for photons compared to Minkowski and Schwarzschild
space-times. However, in the case $\gamma<0$, an increase in the
time delay will happen only on scales for which the term containing
$\alpha^2$ is dominant over the term including $\gamma$.

\section{Limitations on $\gamma$ from experiment}
Let us discuss the possible range and values of the constant
$\gamma$. To do this, we suppose that the value of $\alpha^2$ is
much less than $\gamma$ and hence the term including $\alpha^2$ will
appear only at very large scales such as galactic and cosmological
scales. To determine the range of $\gamma$ we consider some of the
tests performed in our Solar System.

\subsection{Deflection of light rays in the Solar System}In Solar
System tests, the Sun is the central object and we neglect the
effects of other planets. The second term in equation (\ref{32}),
$\delta_{\gamma}=\frac{2\gamma
R}{\left(1-\beta^{2}\right)^{3/2}}\approx 2\gamma R$, is a simple
modification to the standard result, namely
$\delta_{GR}=4\frac{GM_{\odot}}{c^{2}R}$ with $G$, $M_{\odot}$ being
the gravitational constant and the Sun's mass respectively and $c$
is the speed of light. The deflection of a ray that comes from
infinity and grazes the Sun's limb is $\delta_{GR}\approx1.75$
arcsec. To date, the best measurements on the deflection of light
from the Sun have been obtained using very-long-baseline
interferometry data, measuring the deflection of photons emanated
from distant compact radio sources. Since no deviation from general
relativity has been reported \cite{13}, we find that
$|\delta_{\gamma}|\leq2.11\times10^{-10}$ and consequently
$|\gamma|\leq1.52\times10^{-21} cm^{-1}$. Such experiments constrain
the value of $\gamma$, but give no clue as to the sign.
\subsection{Time delay of signals in the Solar System}
According to equation (\ref{6,36}), photons are delayed by the
curvature of space-time characterized by the line element
(\ref{3,13}). To determine the range of $\gamma$, in this section,
we compare the experimental data of time dilation of signals with
predictions afforded by theory. In 1979 \cite{14} the result of
Viking Relativity Experiment confirmed the ``Shapiro'' time delay in
the Solar System to an accuracy of $0.1\%$, but the most recent
result is the frequency shift of radio signals to and from the
Cassini spacecraft as they passed near the Sun \cite{1415}. In
general relativity, the increase in $\Delta t$ produced by the
gravitational field of the Sun over the time taken for photons to
travel the round trip between the ground antenna and the spacecraft
at distances $r_{e}$ and $r_{s}$ respectively from the Sun, is
\cite{14150}
\begin{eqnarray}\label{7,37}
\Delta
t_{_{GR}}=4\frac{GM_{\odot}}{c^3}\ln\left(\frac{4r_{e}r_{s}}{b^2}\right),
\end{eqnarray}
where $G$ is the gravitational constant, $M_{\odot}$ is
gravitational mass of the Sun and $b$ is the impact parameter. It is
convenient to use the relative change in the frequency which is
caused by the gravitational time delay \cite{14151}, because the
Doppler shift due to the receiver's motion has no effect owing to
the cancelation at both the receipt and emission of the radio
signals \cite{14151}. This frequency shift is defined as $y
=-\frac{d(\Delta t )}{dt}$. Indeed, the frequency shift was used by
the Cassini experiment. For a case of $b\ll r_{e}, r_{s}$, which is
valid for the Cassini experiment, the general relativistic
contribution is expressed as \cite{14150}
\begin{eqnarray}\label{7,38}
 y_{_{GR}}=4\frac{M_{\odot}}{b}\frac{db}{dt}.
\end{eqnarray}
We take the extra term caused by $\gamma$ in time delay in equation
(\ref{6,36}) and find the extra frequency shift as
\begin{eqnarray}\label{7,39}
y_{\gamma}=\gamma\left[\frac{r_{0}+2r_{e}}{(r_{0}+r_{e})^2}+\frac{r_{0}+2r_{s}}{(r_{0}+r_{s})^2}\right]b\frac{db}{dt},
\end{eqnarray}
where we have assumed $\frac{dr_{e}}{dt},
\frac{dr_{s}}{dt}\ll\frac{dr_{0}}{dt}\sim\frac{db}{dt}$ near the
solar conjunction ($b\ll r_{e}, r_{s}$). For a spacecraft  much
farther away from the Sun than the Earth, $\frac{db}{dt}$ is not
very different from the velocity of Earth $ v_{e}=30 km s^{-1}$. Now
we can put constraint on $\gamma$ from the Doppler tracking of the
Cassini spacecraft while it was on its way to Saturn reported in
\cite{1415}. From \cite{1415} we find that $|\rm y_{\gamma}|\leq
10^{-14}$ and therefore, at $r_{s}=8.43 AU$, $b_{min}=1.6 R_{\odot}$
we obtain $|\gamma|\leq10^{-28} cm^{-1}$. We gain seven order of
magnitude more accuracy than the constraint from the bending of
light experiments. However, it is still not possible to determine
the sign of $\gamma$.
\subsection{Precession of planetary orbits}
Now, the measured perihelion shift of Mercury is known accurately;
after the perturbing effects of other planets have been accounted
for, the excess shift is known to be about $0.1$ percent from radar
observations of Mercury between 1966 and 1990 \cite{14152,14153}.
The prediction of general relativity for perihelion shift of Mercury
is $\Delta\varphi_{GR}=42.98$ $arcsec/century$ \cite{11,12} while
the observed precession of the perihelion of Mercury is
$\Delta\varphi_{Obs}=43.13 \pm 0.14$ $arcsec/century$ \cite{14154}.
Therefore the difference
$\delta\varphi=\Delta\varphi_{Obs}-\Delta\varphi_{GR}=0.15$
$arcsec/century$ can be used to constrain $\gamma$. From equation
(\ref{4,25}) we obtain the contribution of $\gamma$ term to the
perihelion shift
\begin{eqnarray}\label{8,40}
\Delta\varphi_{\gamma}\approx\gamma\frac{2\pi J^{4}}{m^{3}},
\end{eqnarray}
where we have assumed that $3m\ll\frac{J^{4}}{m^{3}}$ and
$\frac{J^{2}}{m}=a(1-e^{2})$, with $a$ being the semi-major axis and
the eccentricity of planet respectively. By assuming that the
difference $\delta\varphi$ is due to the contribution of the
$\gamma$ term in  metric (\ref{3,13}) (note that  $\alpha^{2}$ is
much smaller than $\gamma$), the observational result imposes the
following constraint on $\gamma$
\begin{eqnarray}\label{8,41}
|\gamma|\leq\frac{GM_{\odot}}{2\pi
c^{2}a^{2}(1-e^{2})}\delta\varphi.
\end{eqnarray}
Use of the observational data for Mercury, equation (\ref{8,41}),
then gives $|\gamma|\leq 1.33\times10^{-30} cm^{-1}$. This result is
two order of magnitude smaller than that given by time delay of
signals.
\subsection{Mean motion}
Due to the extra terms $2\gamma r$ and $\alpha^2r^2$ in the metric,
the radial motion of a test body around a central object will be
affected by an additional acceleration. Let us consider a circular
orbit
\begin{eqnarray}\label{8,42}
r\omega^2=\frac{GM}{r^2}-\gamma c^2-\alpha^2c^2r,
\end{eqnarray}
where $\omega$ is the angular frequency of the orbit. We can rewrite
the last equation as
\begin{eqnarray}\label{8,43}
r\omega^2=\frac{GM_{eff}}{r^2}.
\end{eqnarray}
Here $M_{eff}$ is the effective mass  of the central object which
can affect the motion of the test body. Obviously, a positive
$\gamma$ would decrease the mass of the central object, leading to
an excess in the orbital semi-major axis of the test body. In other
words, the mean motion of the test body $n=\sqrt{\frac{GM}{a^3}}$ is
changed by \cite{17}
\begin{eqnarray}\label{8,44}
\frac{\delta n}{n}=-\frac{1}{2r_{g}}\left(\alpha^2r^3+\gamma
r^2\right),
\end{eqnarray}
where $r_{g}=\frac{GM}{c^2}$ is the gravitational radius of the
central object.

We can evaluate the statistical error on the semi-major axis of each
test body $\delta a=-\frac{2}{3}a\frac{\delta n}{n}$, and interpret
it as the uncertainty in the determination of $\gamma$. Bounds on
$\gamma$ are shown in Table 1 where  $\delta a$ is the statistical
error in the orbital semi-major axis of each planet \cite{18}. We
see that the value of $\gamma$ is between 3 to 5 order of magnitude
smaller than that given by the time delay of signals and about three
order of magnitude smaller than that of the perihelion shift, but
since we are considering a statistical error we have in effect
constrained the absolute value of $\gamma$.

\begin{table}[h]

 \centering
\begin{tabular}{|c|c|c|}
  \hline

    & $\delta a (cm)$ & $(\gamma)_{lim} (cm^{-1})$ \\\hline
 $Mercury$ &10.5 &$
2.38\times 10^{-32}$\\

  \hline $Venus  $ &32.9
& $1.15\times 10^{-32}$\\

 \hline $Earth  $&14.6 &
$1.93\times 10^{-33}$\\

 \hline  $Mars   $&65.7
& $2.46\times 10^{-33}$\\

 \hline $Jupiter$&6390.0 &
$6.00\times 10^{-32}$\\

  \hline $Saturn $&4222.0
&$6.35\times 10^{-32}$\\

 \hline $Uranus $&38484.0 &
$7.15\times 10^{-32}$\\

 \hline $Neptune$&478532.0
&$2.32\times 10^{-31}$\\

  \hline$Pluto  $&3463309.0 &
$7.44\times 10^{-31}$\\

\hline
\end{tabular}
\caption{Limits on $\gamma$ due to anomalous mean motion of planets
in the Solar System. $(\gamma)_{lim}$ is the extreme limit on the
parameter $\gamma$. }\label{TabRadiation}
\end{table}
\section{Conclusions}
In this paper we have studied the classical tests of general
relativity in a brane-world scenario in which the localization of
matter on the brane is through the action of a confining potential.
We have studied the effects of $\alpha^2$ and $\gamma$ parameters on
radar retardation, bending of light rays and the change in the mean
motion of planets and test bodies. These are similar to the familiar
results obtained in a Schwarzschild space-time but with certain
corrections. To determine the value of  constant $\gamma$, we
assumed that the parameter $\alpha^2$ is much less than $\gamma$ and
hence the terms including $\alpha^2$ could be neglected. Using
available data from the Solar System, we obtained different
constraints on $\gamma$. The best constraints come from anomalies in
the mean motion of the Solar System planets resulting in absolute
values for $\gamma$ of $1.93\times 10^{-33} cm^{-1}$ and $2.46\times
10^{-33} cm^{-1}$ for the Earth and Mars respectively.

\end{document}